\documentclass[aip,reprint]{revtex4-1}

\usepackage[english]{babel}
\usepackage{graphicx}
\usepackage{mathptmx}
\usepackage{color}

\newcommand{\pderv}[2]{\frac{\partial #1}{\partial #2}}

\newcommand{\rev}[1]{{#1}}

\begin{document}

\title[Kinetic whistler instability in a continuous  ECR ion source]{Kinetic whistler instability in a mirror-confined plasma of a continuous  ECR ion source}

\author{M. Viktorov}
\email{mikhail.viktorov@ipfran.ru}
\affiliation{Institute of Applied Physics of Russian Academy of Sciences, 46 Ulyanov Street, 603950, Nizhny Novgorod, Russia}
\affiliation{Lobachevsky State University of Nizhny Novgorod, 23 Gagarina Avenue, 603022, Nizhny Novgorod, Russia}

\author{I. Izotov}
\affiliation{Institute of Applied Physics of Russian Academy of Sciences, 46 Ulyanov Street, 603950, Nizhny Novgorod, Russia}
\affiliation{Lobachevsky State University of Nizhny Novgorod, 23 Gagarina Avenue, 603022, Nizhny Novgorod, Russia}

\author{E. Kiseleva}
\affiliation{Institute of Applied Physics of Russian Academy of Sciences, 46 Ulyanov Street, 603950, Nizhny Novgorod, Russia}
\affiliation{Lobachevsky State University of Nizhny Novgorod, 23 Gagarina Avenue, 603022, Nizhny Novgorod, Russia}

\author{A. Polyakov}
\affiliation{Institute of Applied Physics of Russian Academy of Sciences, 46 Ulyanov Street, 603950, Nizhny Novgorod, Russia}
\affiliation{Lobachevsky State University of Nizhny Novgorod, 23 Gagarina Avenue, 603022, Nizhny Novgorod, Russia}

\author{S. Vybin}
\affiliation{Institute of Applied Physics of Russian Academy of Sciences, 46 Ulyanov Street, 603950, Nizhny Novgorod, Russia}
\affiliation{Lobachevsky State University of Nizhny Novgorod, 23 Gagarina Avenue, 603022, Nizhny Novgorod, Russia}

\author{V. Skalyga}
\affiliation{Institute of Applied Physics of Russian Academy of Sciences, 46 Ulyanov Street, 603950, Nizhny Novgorod, Russia}
\affiliation{Lobachevsky State University of Nizhny Novgorod, 23 Gagarina Avenue, 603022, Nizhny Novgorod, Russia}

\date{\today}

\begin{abstract}
Kinetic instabilities in a dense plasma of a continuous ECR discharge in a mirror magnetic trap at the GISMO setup are studied. We experimentally define unstable regimes and corresponding plasma parameters, where the excitation of electromagnetic emission is observed, accompanied by the precipitation of energetic electrons from the magnetic trap. Comprehensive experimental study of the precipitating electron energy distribution and plasma electromagnetic emission spectra, together with theoretical estimates of the cyclotron instability increment proves that under the experimental conditions the observed instability is related to the excitation of whistler-mode waves, which are a driver of losses of energetic electrons from the magnetic trap. The results of this study are important for the further development of the GISMO ECRIS facility and for the improvement of its parameters as an ion source. Also, this research of plasma kinetic instabilities is of fundamental interest and provides experimental tools to simultaneously study plasma electromagnetic activity and corresponding changes in a resonant electron energy distribution.
\end{abstract}

\maketitle

\section{Introduction}


The cyclotron instability of whistler-mode waves is one of the most studied type of plasma kinetic instability realized in space and laboratory magnetic traps \rev{\cite{Ard_1966,Ikegami_1968,Gitomer_1970,Garner_1987,Garner_1990}}. This instability develops in the magnetic tubes filled with dense cold plasma containing usually a small addition of energetic electrons with an anisotropic distribution function \cite{trakhtengerts_rycroft_2008}. Resonant energetic particles can effectively generate whistler waves at frequencies below their gyrofrequency, propagating at small angles to the magnetic field of the trap. As a result of the resonant interaction, energetic particles give off their energy to the waves, fall into the loss-cone in velocity space, and leave the magnetic trap. In many cases, the cyclotron interaction of waves and resonant particles is the only driver of losses of energetic charged particles effectively confined in a magnetic trap.

In the Earth's magnetosphere whistler waves regulate fluxes of trapped electrons and their precipitation rate in the upper atmosphere, determining the wave energy budget in the outer radiation belt \cite{Artemyev2015Nature}. In the nonlinear regime of resonant wave-particle interaction whistler waves can drive precipitation of relativistic radiation belt electrons, also known as microbursts \cite{Chen_2022,Tsai_2022,Artemyev_2022}.

In the laboratory plasma excitation of whistler waves can be triggered by antennas \cite{Stenzel_2015_PRL,Gushchin_2008_PhP,Zudin_2021_JETPL}, by the injection of electron beams into a preexisting plasma \cite{Stenzel_1977,Stenzel_1999,Starodubtsev_1999_PRL,Van_Compernolle_2015,Van_Compernolle_2017}, generated by the unstable electron velocity distribution of a core plasma created under the electron cyclotron resonance (ECR) condition in a mirror magnetic trap \cite{Vodopyanov_2005,Viktorov_EPL_2015,Shalashov17,Viktorov2020} or by runaway electrons in fusion plasmas \cite{Heidbrink_2018,Lvovskiy_2020,Kim_2020,Buratti_2021,Bin_2022}. As well as in the space magnetic traps, in the laboratory devices whistler waves can change the confinement of the energetic electron fraction and cause the precipitation of electrons with energies up to relativistic ones \cite{Vodopyanov_2005,Viktorov2020,Van_Compernolle_2014,Liu_2018}.

The characteristic feature of plasma created and sustained under the ECR conditions in a mirror magnetic trap is the existence of a fraction of energetic electrons with an anisotropic distribution function, which is a source of a free energy for different types of kinetic instabilities \cite{Shalashov17,Mansfeld_DPR_2018,Shalashov_2019,Eliasson20,Shalashov20}. Diagnostics and studies of the kinetic processes in ECR plasma became a very important part in the development of the modern electron cyclotron resonance ion sources (ECRIS) \cite{Izotov_2021_PPCF,Skalyga_2022}.

During the experimental campaign, which lasted over 10 years at JYFL ECRIS (University of Jyvaskyla, Jyvaskyla, Finland), it had been shown that kinetic plasma instabilities in conventional (i.e. low-density high-temperature) ECRIS cause perturbations of the ion current during the plasma decay, whereas the initial transient ion current peak observed during the afterglow has been successfully utilized for injection to circular accelerators of heavy particles \cite{Izotov_2012_PhP}.
In a minimum-B ECRIS plasma sustained by continuous radiation the cyclotron instabilities under certain conditions led to ms-scale oscillation of the extracted beam current, which are especially problematic for modern applications of high performance ECRIS requiring a supreme temporal stability \cite{Tarvainen_2014}. 
The experiments on measuring the microwave emission related to cyclotron instabilities in a minimum-B conventional ECRIS revealed a complex structure of a broadband electromagnetic signals \cite{Izotov_2015_PSST, Izotov_2017_PhP,Bhaskar_2022}. A conclusion had been drawn, assuming that the kinetic instabilities driven by high-energy electrons (tens of keV) restrict the parameter space available for the optimization of high charge state ion currents \cite{Shalashov_2018_PRL, Toivanen_2022}.  

Here, we present the study of plasma kinetic instabilities at the GISMO (Gasdynamic Ion Source for Multipurpose Operation) setup \cite{Skalyga_2022,Skalyga_2019} – ECRIS with plasma confinement in a quasi-gasdynamic regime.
When compared to JYFL ECRIS, GISMO facility has a specific energy input tenfold higher, allowing to sustain the discharge in gasdynamic confinement mode, where the plasma density may reach up to $10^{13}$\,cm$^{-3}$. Such plasma density is inaccessible in previous experiments at the ECRIS/JYFL facility, which makes it possible to study new modes of formation of the electron energy distribution in the trap. We experimentally define regimes, where the excitation of whistler-mode electromagnetic emission is observed, accompanied by the precipitation of energetic electrons from the magnetic trap.

\section{Experimental setup and diagnostics}

\begin{figure*}[tbh]
\includegraphics[width=150mm]{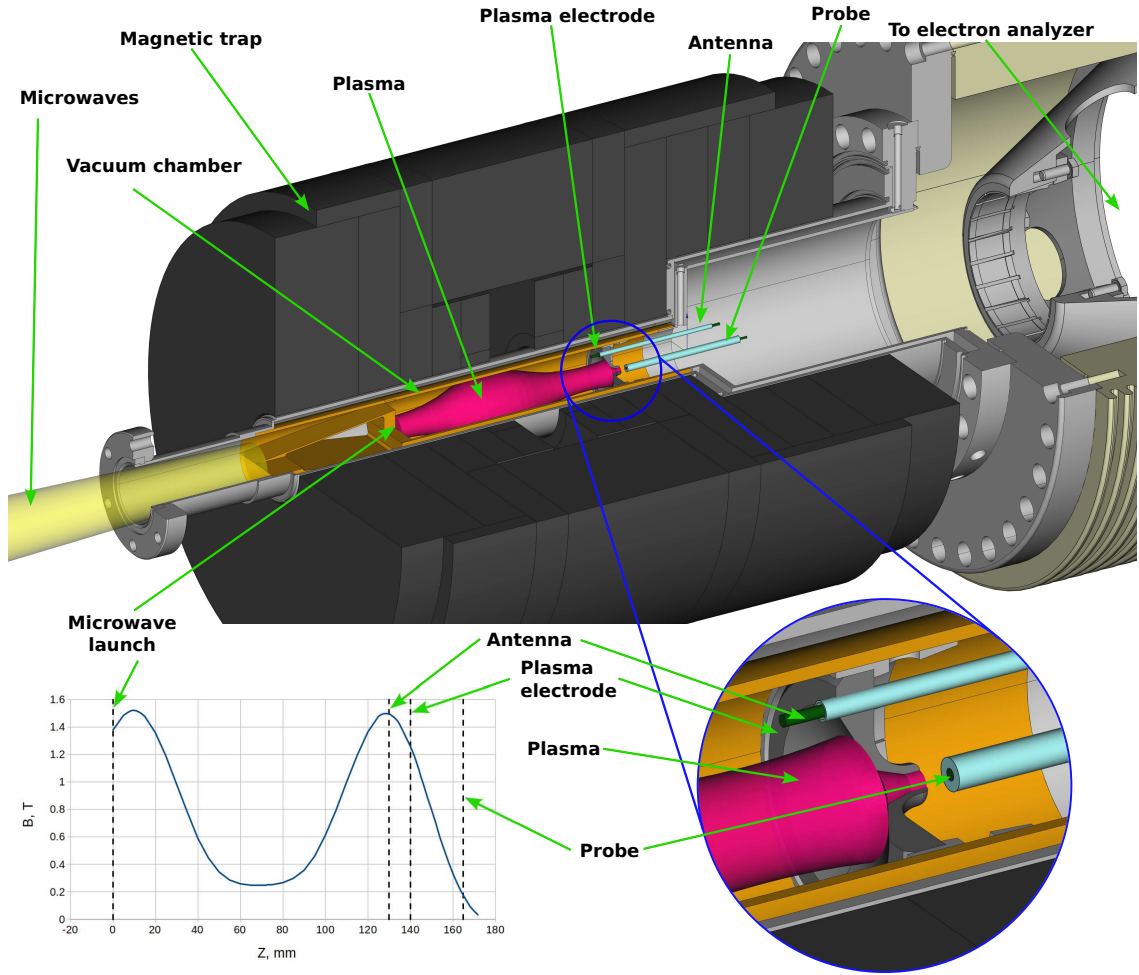}
\caption{\label{fig:setup} A schematic view of the GISMO facility. Microwave radiation is launched along the trap axis and creates plasma (purple volume), which is confined in a mirror magnetic trap. \rev{The insert shows the distribution of the magnetic field strength at the trap axis and corresponding location of the experimental units.}}
\end{figure*}

The experiments were carried out at the GISMO setup, which scheme is shown in Figure~\ref{fig:setup}. For the ECR plasma heating, a modern gyrotron with a continuous-wave power of up to 10 kW at a frequency of 28 GHz was used, which corresponded to a specific energy input of 50–100 W/cm$^3$. 
\rev{In the present experiment, the input heating power was limited to 5\,kW due to the parasitic gas breakdown inside the microwave input unit at a higher power.}
Plasma is confined in a magnetic trap, created by a system of permanent magnets. To ensure sufficient plasma confinement, the magnetic field configuration is designed to be similar to a simple mirror trap close to the system axis, and the magnetic field strength is 1.5 T at plugs and 0.25 T at the trap center, yielding the mirror ratio R = 6. The distance between magnetic mirrors is 12 cm. \rev{The distribution of the magnetic field strength at the trap axis is shown in Figure~\ref{fig:setup}.} The magnet bore is 50 mm, allowing the use of a water-cooled plasma chamber with a vacuum bore of 32 mm. Hydrogen was used as the working gas.

The important advantage of the GISMO facility is a wide range of gas pressures in the discharge (0.001-5 mTorr), which makes it possible to conduct research not only in the classical collisionless, but also in the quasi-gasdynamic (collisional) confinement modes \cite{Skalyga_2022}.

The cold plasma parameters were experimentally measured with a removable Langmuir probe, shown in Figure~\ref{fig:setup}. The probe consisted of a tungsten wire of 1.34\,mm diameter placed inside a ceramic tube. The magnetic field lines were \rev{parallel to the probe surface normal vector}, which makes it possible to apply the theory of a non-magnetized probe, keeping in mind that only the longitudinal electron temperature can be measured. 
The grounded plasma chamber was a reference electrode, whose area is approximately the same as the area of the probe, following the magnetic field lines. Due to comparable areas, the electron density was estimated basing on the ion saturation current, assuming the quasi-neutrality condition. \rev{Electron temperature was estimated under assumptions that electron distribution is Maxwellian.}

To measure the energy distribution of electrons precipitated from the magnetic trap, we used the magneto-static method, which makes it possible to obtain a continuous energy spectrum of electrons with a high temporal resolution \cite{Kiseleva_2022}. \rev{During these measurements the Langmuir probe was removed from the vacuum chamber.} The electron current was detected by a secondary electron multiplier (SEM), located in the vacuum volume after the magnetic analyzer. A voltage of -3.5\,kV relative to the chamber potential was applied to the SEM cathode, which prevented the registration of electrons with energies below 3.5\,keV. The current in the analyzer electromagnet was adjusted by the micro-controller, which allowed scanning with a small energy step. Test calibration measurements proved that the method resolution is about 1\,keV.

\begin{figure*}[tph]
\centering
\includegraphics[width=80mm]{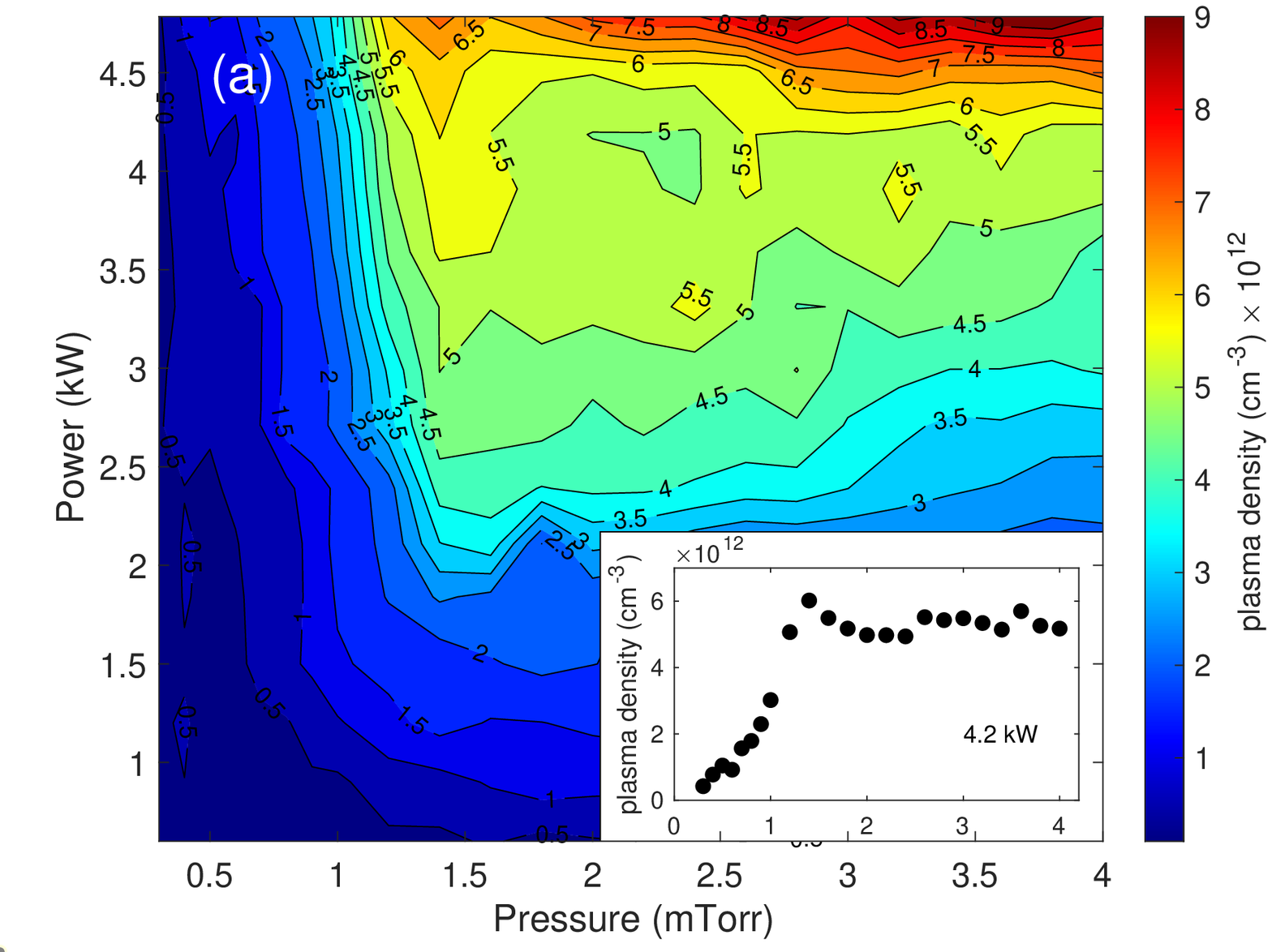}
\includegraphics[width=80mm]{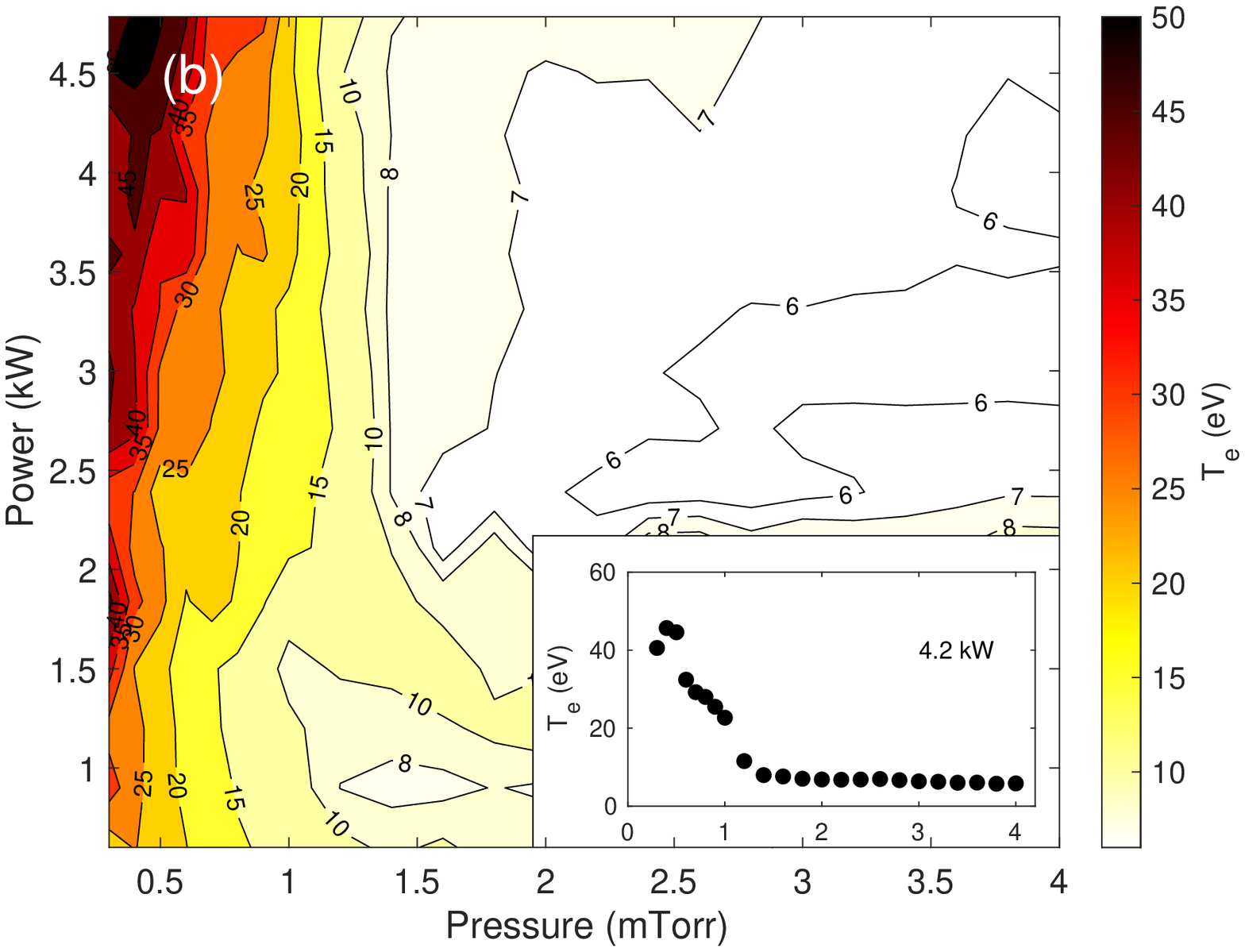}
\caption{\label{fig:plasma_param} (a) Plasma density and (b) the temperature of the main electron fraction obtained from the Langmuir probe measurements. The inserts on both plots correspondingly show the evolution of these values at a fixed heating power of 4.2 kW.}
\end{figure*}

The presence of a magnetic cusp region \rev{(inevitable due to the permanent magnets)} at the GISMO magnetic trap made it possible to observe only electrons moving along field lines located close to the system symmetry axis. A significant part of the electrons flew to the chamber walls, generating bremsstrahlung, and only a small part of the electrons emitted from the plasma moved along the axis of the system, reaching the SEM. When constructing the electron energy distribution, the numerically calculated transport function of the magnetic system was taken into account, which shows what part of the electrons with a given energy will fly through the vacuum path. Also, we take into account the probabilities of electron back-scattering from the SEM cathode and secondary electron emission.

To measure the plasma electromagnetic radiation a coaxial antenna was designed and manufactured. The general view of the antenna mounting scheme is shown in Figure~\ref{fig:setup}. The antenna is a ceramic tube with a diameter of 3\,mm and a length of 50\,mm and a molybdenum core, the end of which is extended from the ceramic tube by 10\,mm in the direction of the plasma. A hole was drilled in the molybdenum plasma electrode at a \rev{radial} distance of 2\,mm from the wall of the plasma chamber, through which the antenna tip is inserted directly into the discharge volume \rev{(see, Figure~\ref{fig:setup})}. The antenna is fixed with a stainless steel clamp, which ensures a tight fit of the ceramic tube wall to the cooled wall of the vacuum chamber outside the discharge volume. The current loop of the antenna closes at the exit of the ceramic tube, where the core is soldered to the core of the vacuum coaxial cable, the screen of which is grounded to the wall of the vacuum chamber.

The antenna frequency response was \rev{approximately estimated numerically by calculating S21 parameter, having the source at the plasma chamber mid-plane and destination at the coaxial cable output. Simulation showed} a flat-top-like shape above 4\,GHz towards higher frequencies and exponential decay towards lower frequency band. It should be noted that despite the rather low coupling coefficient in the low-frequency region, the developed coaxial antenna is capable of detecting radiation over the entire frequency range $2-24$\,GHz, provided that the receiving equipment is sufficiently sensitive. However, in the plasma environment the \rev{cut-off} frequency can be significantly reduced, thus the frequency response of the antenna in reality can turn out to be much stronger, which, apparently, was observed in the experiment, where a significant signal was stably recorded at the frequency 3.5\,GHz.

Electromagnetic radiation received using a coaxial antenna located inside the plasma chamber was recorded using a Keysight DSA-Z 594A broadband digital oscilloscope (analogue bandwidth 59\,GHz, sampling frequency 160\,Gs/s). To protect the receiving path from the powerful radiation of the gyrotron at a frequency of 28\,GHz, a low-pass filter with a cut-off frequency of 24.66\,GHz (at the level of -30\,dB) was used.

\section{Experimental results}

\begin{figure*}[tph]
\centering
\includegraphics[width=150mm]{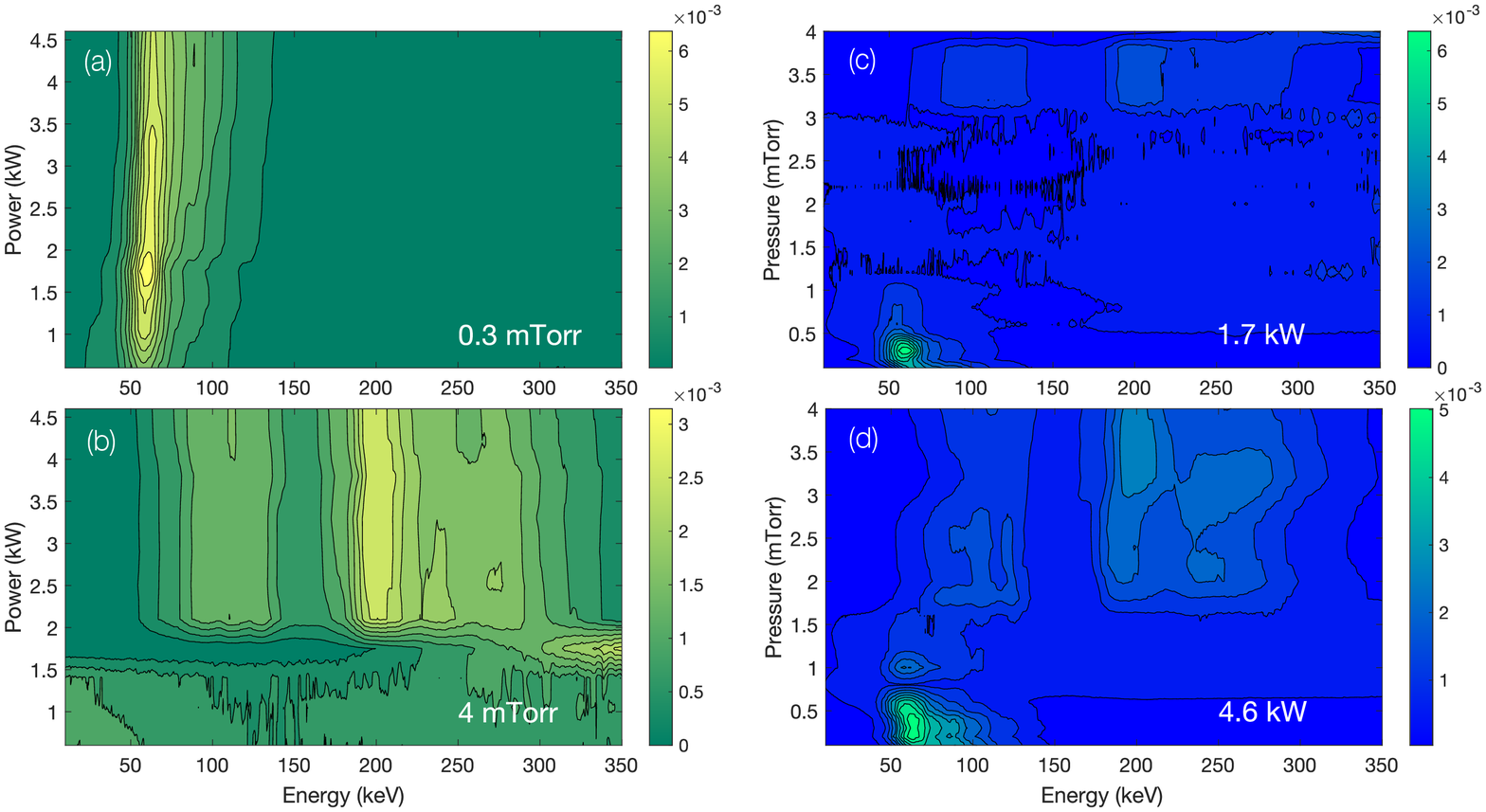}
\caption{\label{fig:eedf_var} \rev{(a)-(b) The precipitated electron energy distribution (EED) at a fixed ambient pressure while heating power is varying and (c)-(d) at a fixed heating power and varying ambient pressure.} The EED is normalized to the unity.}
\end{figure*}

\begin{figure*}[tph]
\centering
\includegraphics[width=80mm]{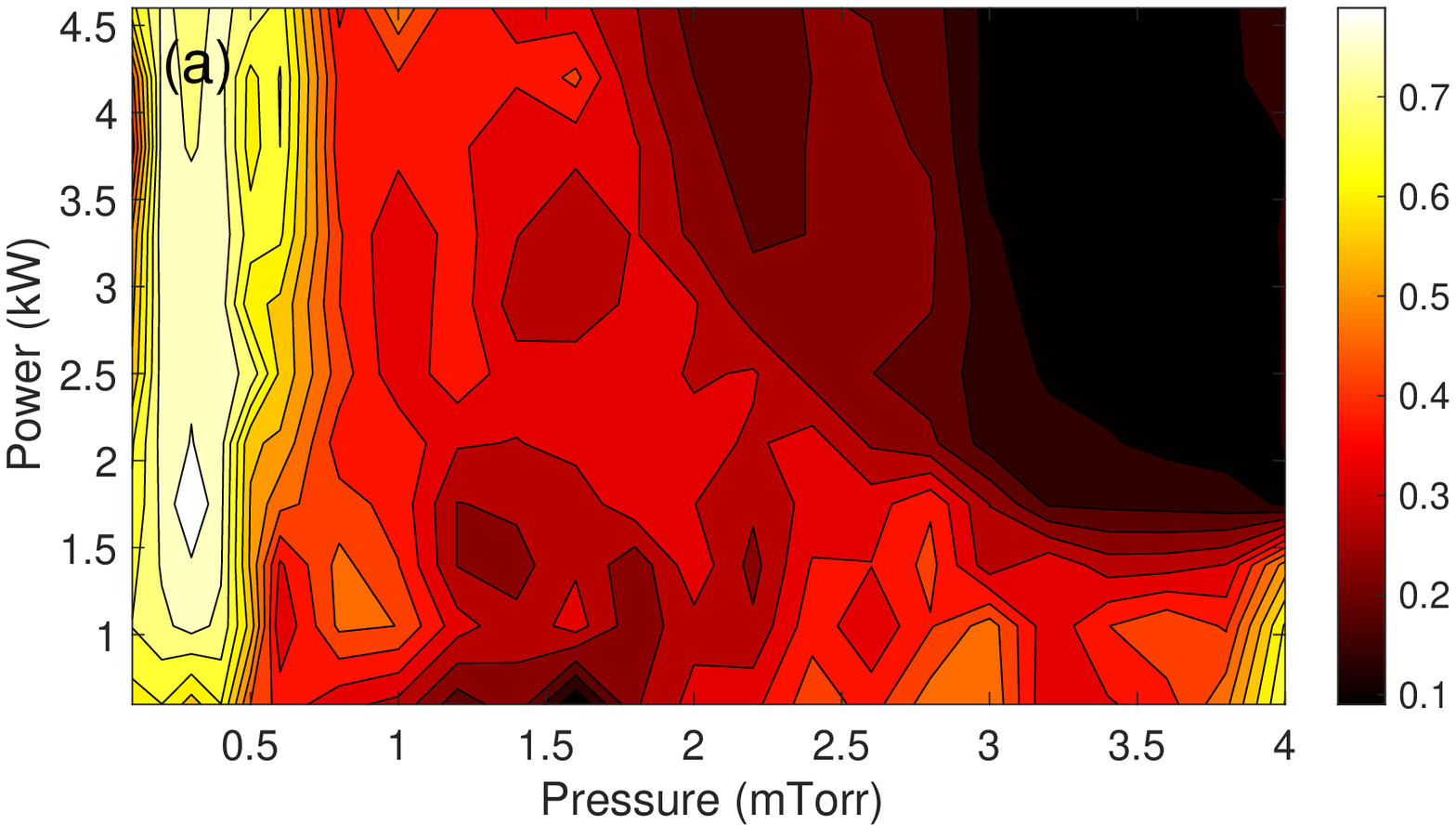}
\includegraphics[width=80mm]{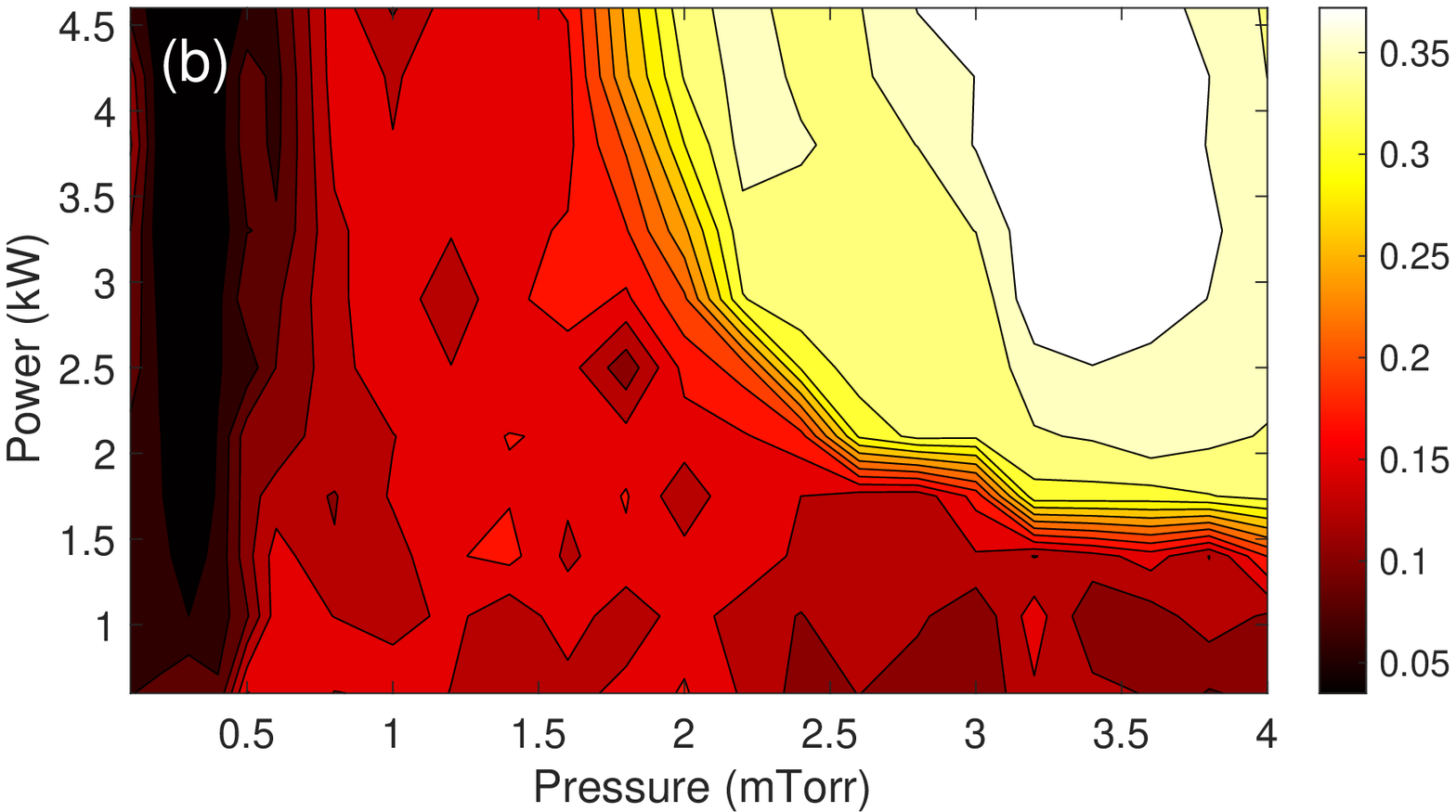}
\caption{\label{fig:eedf_map} The distribution of the EED integral over energies (a) $0-100$\,keV and (b) $180-270$\,keV.}
\end{figure*}

\begin{figure}[th]
\centering
\includegraphics[width=80mm]{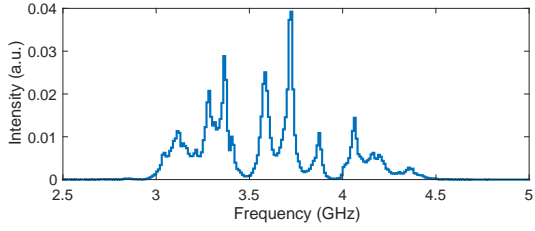}
\caption{\label{fig:freq_hist} Frequency spectrum of plasma electromagnetic emission averaged over the entire dataset.}
\end{figure}

\begin{figure*}[tbph]
\centering
\includegraphics[width=80mm]{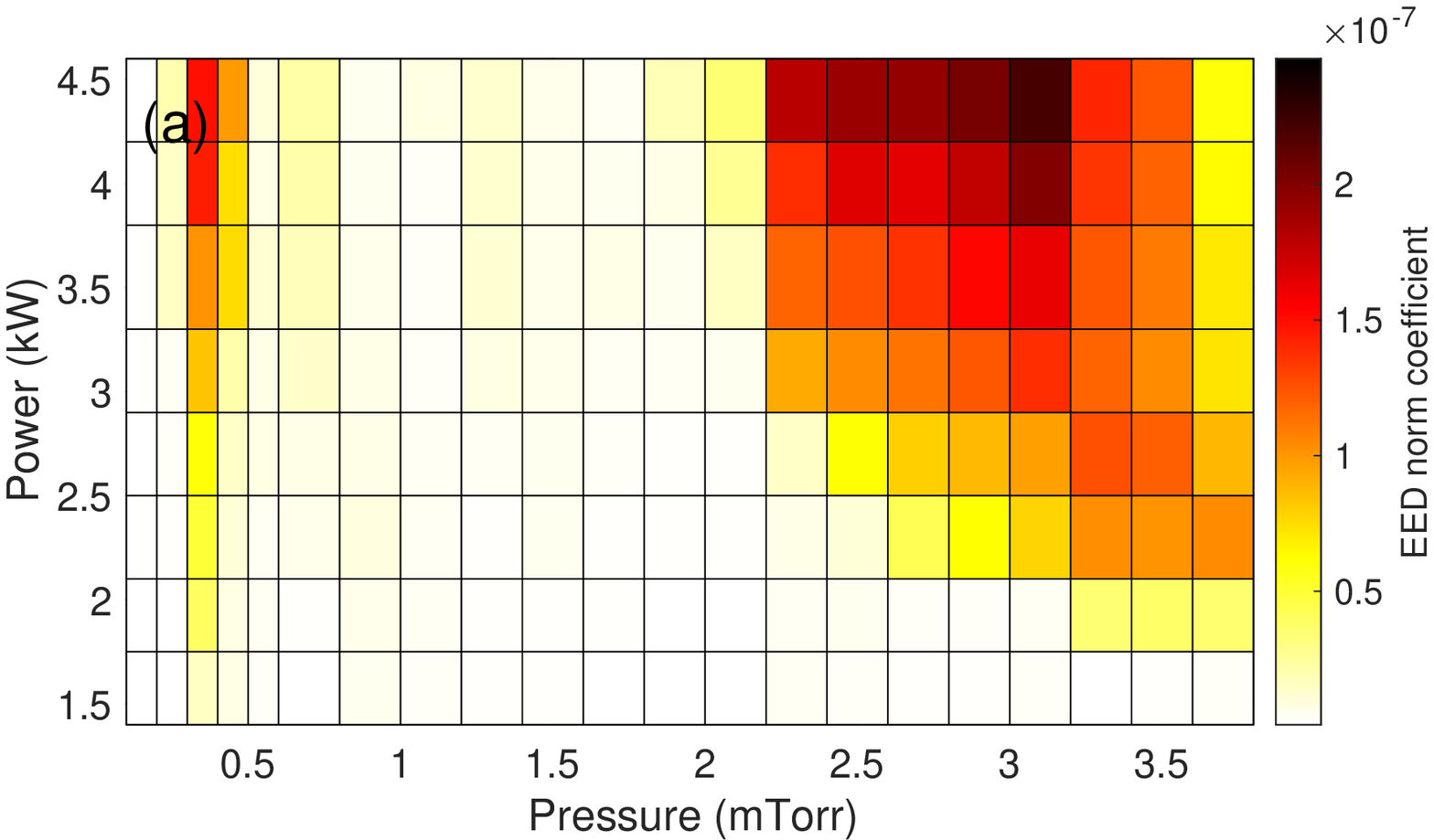}
\includegraphics[width=80mm]{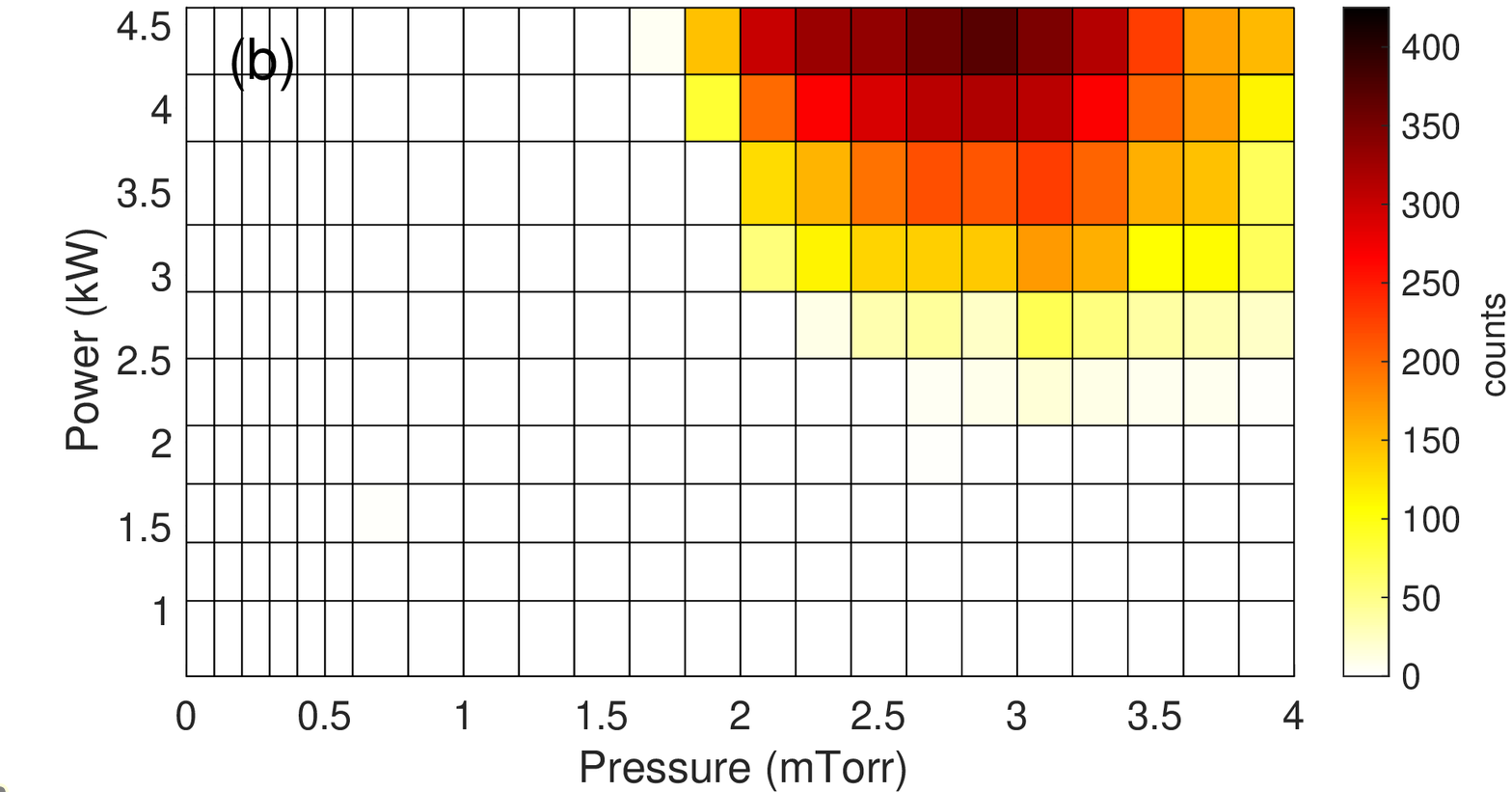}
\caption{\label{fig:bursts} (a) The EED normalization coefficient and (b) the number of recorded electromagnetic bursts (waveform duration is 2\,ms).}
\end{figure*}


The objective of a current research is to understand the development of plasma cyclotron instabilities resulting in the excitation of electromagnetic emission, accompanied by the precipitation of energetic electrons from the magnetic trap. The studies were carried out in a hydrogen plasma sustained by a continuous gyrotron radiation with a power in the range of $0.6–4.6$\,kW at an ambient pressure from 0.1 to 4\,mTorr. For each point from the set of parameters we measured plasma density and the electron temperature using the Langmuir probe. Then we simultaneously recorded the waveform of electric field oscillations received from the coaxial antenna with a duration of 2 ms and the current of precipitated electrons from SEM during energy scan with a duration of 50\,s. 

Results of the Langmuir probe measurements are shown in Figure~\ref{fig:plasma_param}. Plasma density is increasing with the increase of both ambient pressure and heating power. We observe significant increase of plasma density for pressures greater than 1.2\,mTorr and heating power greater than 2.5\,kW. At the maximum heating power of 4.6\,kW plasma density is up to $9\times 10^{12}$\,cm$^{-3}$, which is slightly below the value of the critical plasma density $9.7\times 10^{12}$\,cm$^{-3}$ for the heating frequency 28\,GHz. In contrast, the electron temperature is \rev{higher} at small pressures below 1.2\,mTorr and reaches the value of 50\,eV at a maximum power. In the region of a dense plasma the electron temperature is about 6\,eV. Throughout all measurements the temperature approximation error was negligible, thus the total measurement error was estimated by data dispersion to be on the level of 30\%.

The electron energy distribution (EED) was calculated from the current of precipitated electrons using an algorithm described in Ref.~\onlinecite{Kiseleva_2022}. The EED for a fixed ambient pressure, while the heating power is varying is shown in Figure~\ref{fig:eedf_var} at left column, and similarly for a fixed heating power and varying ambient pressure in Figure~\ref{fig:eedf_var} at right column. At all plots the EED is normalized to the unity. 
At a low pressures below 1\,mTorr there is only one fraction of precipitating electrons with the energy $50-60$\,keV. With the increase of an ambient gas pressure to 2\,mTorr and a heating power to 1.7\,kW more energetic electron fractions appear at energies $80-120$\,keV, $180-220$\,keV and $240-270$\,keV. Figure~\ref{fig:eedf_map} shows the distribution of the EED integral over energies $0-100$\,keV and $180-270$\,keV, which supports the fact that these energy fractions in the flow of precipitated electrons originates at different discharge parameters.

To measure the electromagnetic emission bandwidth, we analyzed 268 experimental shots at different heating power and ambient pressure of hydrogen. Windowed Fourier transform was applied to every recorded waveform and then spectrograms were filtered using a noise reduction algorithm. As a result, we obtained a set of time slots where only intense emissions are present. Based on the analysis of the entire dataset, plasma electromagnetic emission is present at frequencies 3-4.5\,GHz, see Figure~\ref{fig:freq_hist}. The spectrum is rather complicated and is similar to those, observed previously at the SMIS37 facility - pulsed gasdynamic ECR ion source \cite{Viktorov2020}. \rev{The details of the spectrum fine structure and its dependency over the experimental parameters will be analyzed in a future work.}

The level of plasma instability at certain parameters of the setup could be characterized by the total current of the precipitated electrons, which is proportional to the EED normalization coefficient, and by the number of recorded electromagnetic bursts (or repetition rate), shown in Figure~\ref{fig:bursts}. It is seen, that bursts of electromagnetic emission are observed when ambient pressure is in the region 1.7-4 mTorr and heating power is greater than 2.5 kW. The same parameter set corresponds to a region with a high current of precipitated electrons. The maximum number of instability events is observed at a pressure 2.8 mTorr and a maximum power of 4.5 kW. However, at low pressures 0.3-0.5 mTorr there is another maximum in the current of precipitated electrons with energies about 50\,keV, while the electromagnetic emission is not observed.

\section{Discussion}


The experimental data clearly shows the existence of two significantly different regimes of the ECR discharge with respect to the injected microwave heating power and ambient pressure of a neutral gas. The first regime appears at a low pressures below 1.2\,mTorr or at a low heating power below 1.7\,kW (let's call this region of discharge parameters - region (I)). Under these conditions plasma density is about $10^{12}$\,cm$^{-3}$ or less, and the electron temperature of a bulk plasma is 20-50\,eV. In the region (I), we observe precipitation of energetic electrons from the magnetic trap with a mean energy 50-60 keV \rev{(see, Figure~\ref{fig:eedf_var}(a) and \ref{fig:eedf_var}(c))}. The form of the bulk plasma electron temperature distribution, shown in Figure~\ref{fig:plasma_param}(b), coincides with the location of the maximum in the current of precipitated electrons with energies up to 100\,keV, see Figure~\ref{fig:eedf_map}(a). In this regime, plasma electromagnetic emission has never been registered.

The second regime occurs at ambient pressure greater than 2\,mTorr and heating power higher 1.7\,kW (region (II)). Under these parameters plasma density is about $(4-6)\times 10^{12}$\,cm$^{-3}$ and reaches $9\times 10^{12}$\,cm$^{-3}$ at a maximum heating power of 4.6\,kW. The plasma electron temperature is about 6\,eV. Here, most of the energy of precipitated electrons is concentrated in two energy fractions: $180-220$\,keV and $240-270$\,keV \rev{(see, Figure~\ref{fig:eedf_var}(b) and \ref{fig:eedf_var}(d))}. The region of a maximum current of precipitated electrons with energies $180-270$\,keV, shown in Figure~\ref{fig:eedf_map}(b), coincides with the parameter area (II). At the same time, under this regime we observe plasma electromagnetic emission as a series of electromagnetic bursts at frequencies $f=3-4.5$\,GHz.  

So, we may define two regions of plasma parameters: (I) less dense plasma with more energetic main electron component and (II) much denser and colder plasma. Electromagnetic emission from the plasma accompanied by precipitation of energetic electrons is observed only in the region (II), which is most naturally related to the development of plasma kinetic instabilities.
\rev{The absence of plasma electromagnetic emissions in the parameter region (I) suggests that here a stable velocity distribution of electrons with energies about 60\,keV is formed, and registered current of such electrons from the magnetic trap is due to the Coulomb or rf-scattering. }

\rev{Plasma dispersive properties are defined by the cold bulk plasma fraction with electron energies 5-50\,eV (for different regimes) with isotropic electron velocity distribution and density $(1-9)\times 10^{12}$\,cm$^{-3}$. Whereas, excitation of electromagnetic waves is determined by the energetic fraction of electrons with energies $50-300$\,keV and anisotropic velocity distribution and much less number density. Typically, in such ECR discharges the ratio of energetic to cold plasma density is about $n_h/n_c \approx 0.01$ \cite{Vodopyanov_1999}.}

The design of a magnetic system shows that the minimum magnetic field is located at the center of the magnetic trap and equals to 0.25\,T, which corresponds to the electron cyclotron frequency $f_{ce0}=7$\,GHz. The observed plasma emission frequency is always below the value of $f_{ce0}$.
The electromagnetic emission of dense plasma \rev{(electron plasma frequency is much high than electron gyrofrequency)} at frequencies below the electron cyclotron frequency is related to the cyclotron instability of whistler waves. At large densities of the background plasma, the cyclotron instabilities of the extraordinary waves are suppressed because their dispersive properties are strongly modified by the background plasma.  
The development of the electron cyclotron instability causes the decrease of a pitch-angle of a resonant electrons and its subsequent shift to the loss-cone, which explains the observed current of energetic electrons in the region (II).

\rev{Another type of plasma waves could be excited in the magnetoactive dense plasma -- the electron Bernstein waves, which are electrostatic waves at the harmonics of the electron cyclotron frequency. Such plasma waves were recently detected in a similar experiment at a pulsed ECR ion source SMIS\,37 at frequencies in the range 1–5 $f_{ce}$. \cite{Eliasson20}. The emissions are attributed to electrostatic instabilities involving a warm 200–400\,eV mirror-confined population of electrons with a ring-like velocity distribution and a much colder (about 1\,eV) component confined by the positive plasma potential, while densities of these fractions are comparable. In the present experiment we do not observe such plasma emissions, which could be due to the absence of the warm electron component.}

To understand the development of plasma instabilities we calculate the linear growth-rate $\gamma\equiv\mathrm{Im}\,\omega$ of the whistler wave instability \cite{trakhtengerts_rycroft_2008} for the non-relativistic regime under the assumption of a homogeneous plasma as:
\begin{eqnarray}
\label{eq:inc_gen}
\gamma=&\frac{2\pi^3e^2}{m_\mathrm{e}c\: \partial(N_{||}\omega)/\partial\omega}\int \left(\pderv{F}{\upsilon_{||}}+\frac{\omega_{\mathrm{ce}}}{k_{||}\upsilon_\perp}\pderv{F}{\upsilon_\perp}\right) \nonumber\\
&\times\:\delta(\omega-k_{||}\upsilon_{||}-\omega_{\mathrm{ce}})\:\upsilon_\perp^3\mathrm{d}\upsilon_\perp\mathrm{d}\upsilon_{||},
\end{eqnarray}
where $F(\upsilon_{||},\upsilon_\perp)$ is the \rev{energetic} electron distribution function  over parallel and perpendicular velocities to the external magnetic field normalized over energetic particle density $n_h$, $\delta$-function defines the cyclotron resonance  condition, $\omega_{ce}=eB/m_e c$ is the electron cyclotron frequency, $\omega=2\pi f$ is a wave cycle frequency, $k_{\parallel}=N_{||}\omega/c$ is a wave-vector component along the  magnetic field, $N_{||}$ is a corresponding refractive index, $m_e$ and $e$ are respectively electron mass and charge, and $c$ is the speed of light. Hereafter we will use approximations for a refractive index $N_{||}\approx\omega_\mathrm{pe}/\sqrt{\omega(\omega_\mathrm{ce}-\omega)}$ and a wave group velocity $\upsilon_\mathrm{gr}=\partial\omega/\partial k_{||}\approx 2c\,(1-\omega/\omega_\mathrm{ce})^{3/2}\sqrt{\omega\omega_\mathrm{ce}}/\omega_\mathrm{pe}$ valid for the whistler wave propagating strictly parallel to the magnetic field, where $\omega_{pe}=\sqrt{4\pi n_c e^2/m_e}$ is the electron plasma frequency and $n_c$ is a cold plasma density. 

\begin{figure}[tbph]
\centering
\includegraphics[width=80mm]{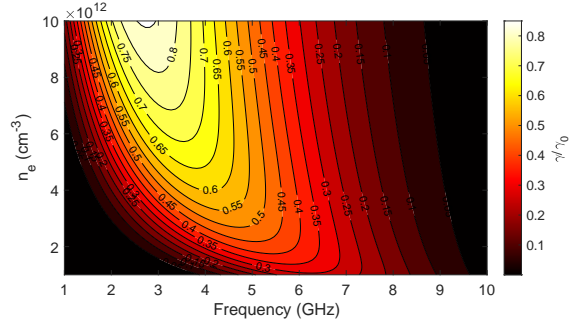}
\caption{\label{fig:inc_map} The increment of the whistler wave instability at different frequencies and density of a cold core plasma. $T_{||}=200$\,keV, $\alpha=T_{\perp}/T_{||}=5$, $f_{ce}=14$\,GHz.}
\end{figure}

In the first approximation, the distribution function $F(\upsilon_{||},\upsilon_\perp)$ of accelerated by the ECR heating electrons may by modeled in a  weakly relativistic limit by bi-Maxwellian distribution:
\begin{equation}
\label{eq:eq_edf} 
F=n_h\frac{m_e^{3/2}}{(2\pi)^{3/2}T_{\perp}T_{||}^{1/2}}\exp\left(-\frac{m_{\mathrm{e}} \upsilon_{||}^2}{2T_{||}}-\frac{m_{\mathrm{e}} \upsilon_{\perp}^2}{2T_{\perp}}\right),
\end{equation}
where $T_{||}$ and $T_{\perp}$ are mean energies of electrons with respect to the external magnetic field.
Substituting (\ref{eq:eq_edf}) into (\ref{eq:inc_gen}) one may obtain,
\begin{equation}
\label{eq:inc}
\frac{\gamma}{\gamma_0} = \frac{\sqrt{\pi}}{2}\alpha \frac{(1-\widetilde{\omega})^{5/2}}{\sqrt{\widetilde{\omega} \beta_*}}
\left(\left(1-\alpha^{-1}\right)-\widetilde{\omega}\right)\exp\left({-\frac{(1-\widetilde{\omega})^3}{\widetilde{\omega} \beta_*}}\right),
\end{equation}
where $\gamma_0=\omega_{ce} n_h/n_c$, $\widetilde{\omega}=\omega/\omega_{ce}$, $\alpha=T_{\perp}/T_{||}$ is the distribution function anisotropy index, $\beta_*=(\omega_{pe}/\omega_{ce})^2 2T_{||}/ m_e c^2$.
\rev{This approach is valid even for resonant electrons with relatively high energies, such as in our research. The only valuable relativistic effect in the equation (\ref{eq:inc_gen}) is the inclusion of the relativistic electron cyclotron frequency in the electron cyclotron resonance condition. But for the case of whistler waves propagating parallel to the magnetic field this correction is negligible due to the high refractive index of the excited waves.}

\begin{figure}[tbph]
\centering
\includegraphics[width=80mm]{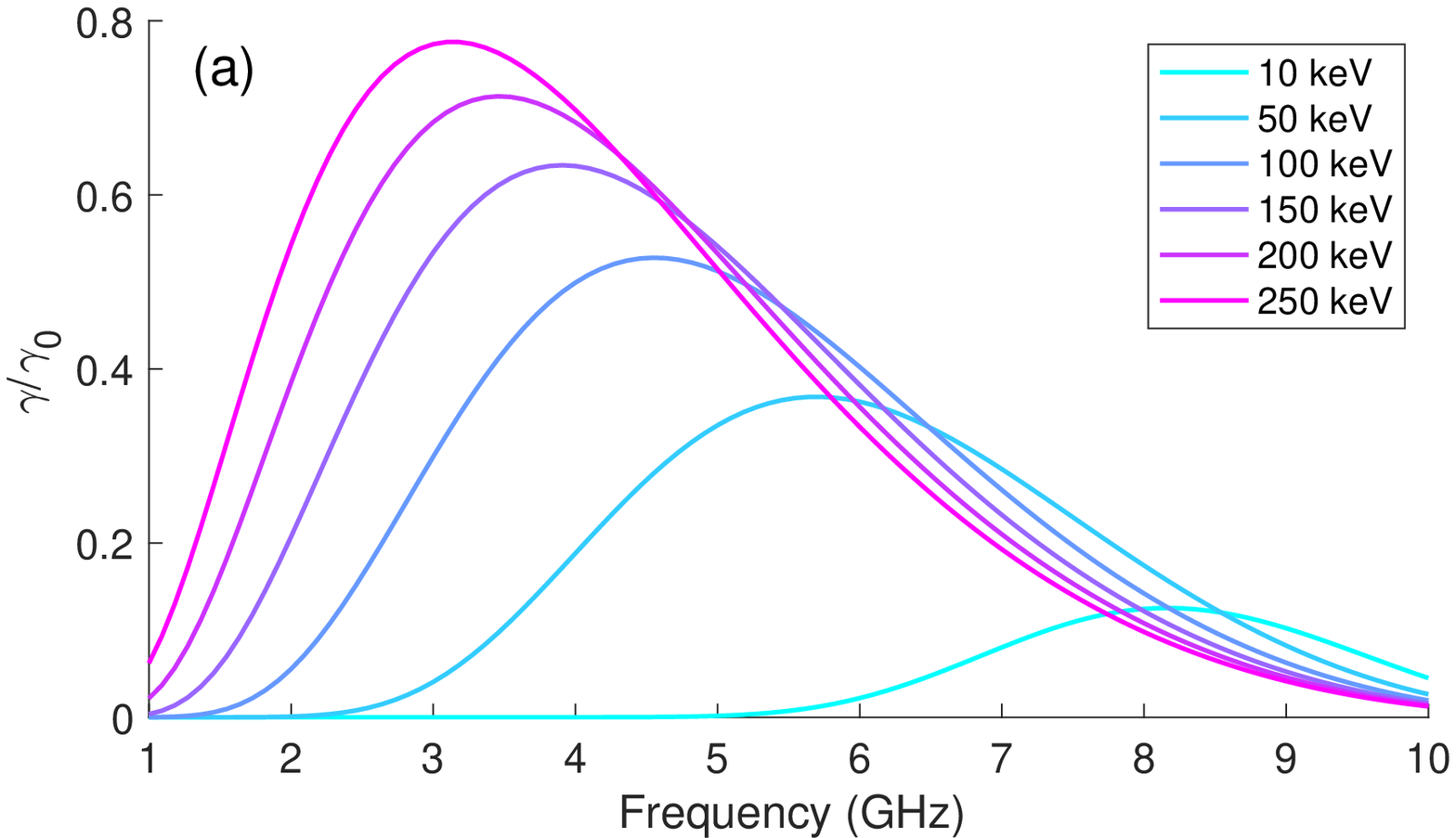}\\
\includegraphics[width=80mm]{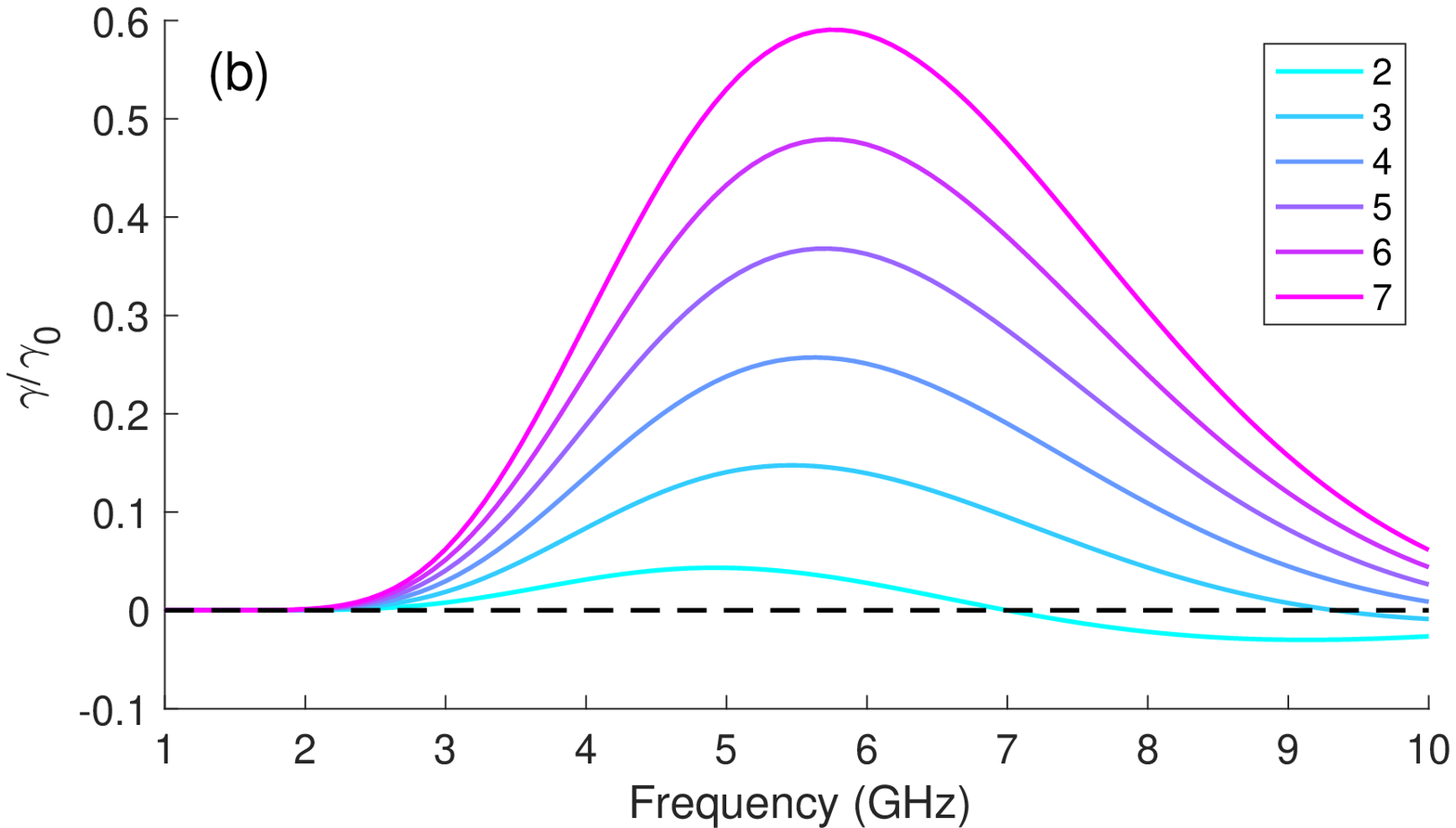}
\caption{\label{fig:inc_Te_alpha} The evolution of the whistler wave instability increment with the charge of (a) mean energy $T_{||}$ of energetic electron fraction ($\alpha=5$) and (b) the distribution function anisotropy index $\alpha$ ($T_{||}=50$\,keV).}
\end{figure}

\begin{figure}[tbph]
\centering
\includegraphics[width=80mm]{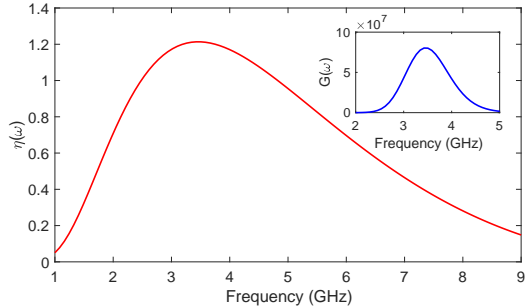}
\caption{\label{fig:gain} The whistler wave amplification coefficient and the amplification factor for the multi-pass amplification ($q=15$), shown in the insert. $n_c=6\times 10^{12}$\,cm$^{-3}$, $n_h/n_c=0.01$, $T_{||}=200$\,keV, $\alpha=T_{\perp}/T_{||}=5$, $f_{ce}=14$\,GHz, $L=12$\,cm.}
\end{figure}

Using (\ref{eq:inc}), we compute the increment of the whistler wave instability for the parameters, relevant to the experiment. Figure~\ref{fig:inc_map} shows the increment values at different frequencies and density of a cold core plasma, calculated for $T_{||}=200$\,keV, which is characteristic for the experimentally observed electron precipitations in region (II). 
\rev{Results are calculated for a value of the electron cyclotron frequency in the source region $f_{ce}=14$\,GHz, which corresponds to the middle point in the magnetic trap between the heating ECR zone and the trap center.}
It is seen, that the unstable frequency band corresponds to the frequency band of experimentally registered plasma emission and the maximum of the increment moves to the lower frequencies with the increase of plasma density. The increase of the energy of a hot electron fraction also leads to the shift of unstable region to the lower frequencies together with the increase of the increment absolute value, see Figure~\ref{fig:inc_Te_alpha}(a). However, the change in the anisotropy index $\alpha$ does not affect the band of instability, see Figure~\ref{fig:inc_Te_alpha}(b).

Following Ref.~\onlinecite{Viktorov2020}, we may define the local spatial amplification coefficient $\eta\equiv 2\mathrm{Im}\,k_{||}$ and the amplification gain $\Gamma$ for a single passage of the wave from one end of the magnetic trap to another as 
\begin{equation}\label{eq:gain_int}
\eta(\omega,z)=2\gamma/\upsilon_\mathrm{gr},\quad 
\Gamma(\omega) = \int\displaylimits_{-L/2}^{L/2}\eta(\omega,z)\mathrm{d}z,
\end{equation}
where $L=12$\,cm is a length of the magnetic trap.
To estimate the amplification gain we assume that plasma parameters are homogeneous along the length of the magnetic trap and magnetic field is constant, thus $\Gamma(\omega) \approx \eta(\omega)L$. The excited electromagnetic wave may pass the resonance region several times due to the reflection from the metal walls of the discharge chamber. The total amplification factor for the multi-pass amplification is $G(\omega)=\exp(q\Gamma(\omega))$, where $q$ is a number of passes through the plasma volume. Computation of the wave amplification, shown in Figure~\ref{fig:gain}, proves that the bandwidth of excited whistler waves with $q=15$ is similar to the experimentally observed data. In reality, the number of passes could be much higher since the Q-factor of the discharge chamber is about 1000.

More precise theoretical analysis of cyclotron instability could be done (similar to Refs.~\onlinecite{Viktorov2020,Shalashov_2021_whist}), taking into account the variation of the magnetic field \rev{(and, correspondingly, the electron cyclotron frequency)} along the trap axis and the details of the energetic electron energy distribution function, such as the loss-cone and EED dependency over the magnetic field.  This work will be published elsewhere. However, already performed analysis shows a good agreement with the experimental data.

\section{Summary}

Kinetic instabilities in a dense plasma of a continuous ECR discharge in a mirror magnetic trap at the GISMO setup are studied. We experimentally define unstable regimes and corresponding plasma parameters, where the excitation of electromagnetic emission is observed, accompanied by the precipitation of energetic electrons from the magnetic trap. Comprehensive experimental study of the precipitating electron energy distribution and plasma electromagnetic emission spectra, together with theoretical estimates of the cyclotron instability increment proves that under the experimental conditions the observed instability is related to the excitation of whistler-mode waves. 

With respect to the previous studies of unstable regimes at other classical ECRIS facilities (such as JYFL ECRIS \cite{Izotov_2015_PSST,Izotov_2017_PhP}), here we define the type of unstable plasma mode, which is a driver of losses of energetic electrons from the magnetic trap. Also, in contrast to the experiments in Ref.~\onlinecite{Izotov_2017_PhP}, we observe electromagnetic emission only below 4.5\,GHz and no emission is ever registered in a higher frequency domain. The analysis of the fine structure of the electromagnetic emission spectrum will be done in a future work.

The results of this study are important for the further development of the gasdynamic GISMO ECRIS facility and will be used for the improvement of its parameters as an ion source. Specific range of plasma parameters in the unstable regime brings new challenges in the implementation of ECRIS stabilization techniques, such as two-frequency heating \cite{Skalyga_2015_PhP}.

\begin{acknowledgments}
The work has been supported by the Russian Science Foundation (grant No.~21--12--00262).
\end{acknowledgments}

\bibliography{refs}

\end{document}